\documentstyle[11pt,paspconf,epsf]{article}

\begin{document}

\title{AGN Variability}

%\author{S. Djorgovski\altaffilmark{1,2} and Ivan R. King}
\author{Toshihiro Kawaguchi and Shin Mineshige}
\affil{Department of Astronomy, Kyoto University, Sakyo-ku, \\
	Kyoto 606-8502, Japan}

\begin{abstract}
Number of monitoring observations of continuum emission from Active Galactic 
Nuclei (AGNs) have been made in optical--X-ray bands. 
%The results obtained so far show the following three features.
%(i) There is no typical timescale in light curves over years to days/hours. 
%(ii) Amplitudes of the variations in those timescales
%are decreasing as the timescales get shorter.
%(iii) Light curves on timescales longer than decades are qualitatively 
%different from ones on shorter timescales. 
%The formers seem to just fluctuate nearly randomly.
The results obtained so far show 
(i) random up and down on timescales longer than decades, 
(ii) no typical timescales of variability on shorter timescales and 
(iii) decreasing amplitudes as timescales become shorter.
The second feature indicates that any successful model must produce 
a wide variety of shot-amplitudes and -durations over a few orders 
in their light curves. 
In this sense, we conclude that the disk instability model is 
favored over the starburst model, 
since fluctuations on days are hard to produce by the latter model. 

Inter-band correlations and time lags also impose great constraints on models.
Thus, constructing wavelength and time dependent models 
remains as a future work.
\end{abstract}

% Keywords should be included, but they are not printed in the hardcopy.


% That's it for the front matter.  On to the main body of the paper.
% We'll only put in tutorial remarks at the beginning of each section
% so you can see entire sections together.

\section{Introduction}

Emission from Active Galactic Nuclei (AGNs) shows 
rapid and apparently random variability 
over wide wavelength ranges from radio to X ray or $\gamma$ ray
(for a review, see Ulrich, Maraschi \& Urry 1997).  
% Krolik et al. 1991; Edelson et al. 1996; 
In spite of numerous intensive, multi-wavelength monitoring projects,
there still remain major questions; 
what causes the variability? 
Which parts of nuclei are emitting in various energy bands? 
One of the goals of variability studies is
to identify and characterize the physical processes responsible 
for the observed variability.

%One of the most important features is, in short, absence of any 
%apparent periodicity or typical timescale. 

Most of AGN light curves exhibit neither apparent periodicity 
nor typical timescale. 
In fact, the AGN light curves fluctuate over wide timescales 
(e.g. Fahlman \& Ulrych 1975, for decades' light curve; 
Peterson et al. 1994, for years; Clavel et al. 1991, for months; 
Korista et al. 1995, for weeks; Edelson et al. 1996, for days). 
Figure 1 shows schematic light curves of AGNs 
on various timescales, from decades to days/hours. 
\begin{figure}
\plotfiddle{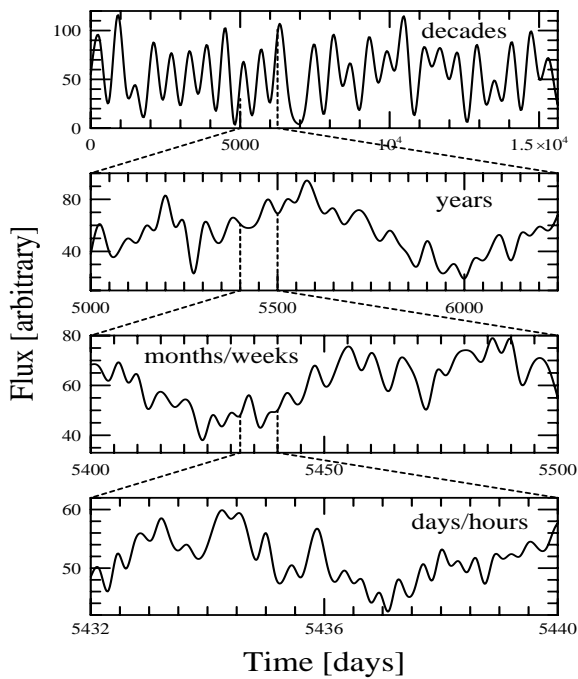}{90mm}{0}{40}{34}{-130}{-10}
%\plotone{lc.epsi}
\caption{Schematic light curves of AGNs 
on different timescales.} 
\end{figure}
We point out following three features;
\begin{enumerate}
\item The top panel, light curves on decades, is qualitatively 
different from the lower three ones. It seems to just fluctuate almost 
randomly, up and down.
\item There is no typical timescale in variability over less than years 
(see the lower three light curves). 
Flux variations over years to days/hours resemble each other; 
each light curve shows one or two bumps 
and small fluctuations superposed on them.
In other words, the variations are self-similar or fractal. 
\item Amplitudes of the variations over less than years 
decrease as the timescales become shorter.
\end{enumerate}

To understand these features above in a statistical way, 
we often use the power-spectral density (PSD) and 
the structure function (SF). 
Observed PSD generally shows a power-law decline 
which is proportional to $f^{-\alpha}$. 
That corresponds to the fact that most of AGN light curves 
show no typical timescale and these amplitudes of the variations 
decrease as the timescales become shorter. 
The power-law index ($\alpha$) ranges from 1.5 to 2.0 
(e.g. Leighly \& O'Brien 1997; Hayashida et al. 1998).
This index includes valuable information and useful 
to discriminate possible theoretical models.
Generally, SF also has a power-law index, $\beta$, 
which is related to $\alpha$ by $\alpha = 1+2\beta$ (\S 2.1). 
On the other hand, when we analyze long light curves, 
we find one critical frequency ($f_{\rm break}$) under which PSD flattens 
and one timescale ($\tau_{\rm var}$) over which SF flattens. 
These frequency and timescale are related each other 
[$\tau_{\rm var} \approx (2 \pi f_{\rm break})^{-1}$].
We often call $\tau_{\rm var}$ as a (maximum) variability timescale, 
and typically it is around 2 - 3 years (e.g. Hook et al. 1994).

Then, the main question in this paper arises; 
\begin{itemize}
\item {\bf What do the variability timescale ($\tau_{\rm var}$) 
and the power-law index of PSD ($\alpha$) or that of SF ($\beta$) tell us?}
\end{itemize}

In Section 2, we focus on the question above, 
presenting our time series analysis of the light curves. 
There, we introduce the two possible models for AGN optical variability: 
the disk instability model and the starburst model, 
and show how we can interpret the observed fluctuation properties 
based on the two models. 
Other observed properties, such as wavelength-dependency of 
the AGN variability, are discussed in Section 3. 
Finally, we summarize the conclusion and address issues 
remaining as future works.

\section{Comparison of Models}

In this section, we are analyzing light curves obtained by 
the observation, disk instability (DI) model (Mineshige et al. 1994), 
and starburst (SB) model (Aretxaga et al. 1997). 
The main issues are the variability timescale 
and power-law index of SF of AGN light curves (Kawaguchi et al. 1998). 

\subsection{Observed Fluctuation Properties} \label{observation}

%In this section, 
%since multi-wavelength monitoring projects show that flux variations 
%in different wavebands from Optical to X-rays are almost coherent, 
%we do not intend to specify our analysis to any wavebands.

Light curves ideal for our analysis are 
long-term observational data over years, which have high sampling 
rates and good photometric accuracy. 
According to these criteria, we chose 
the optical light curve of the double quasar 
0957+561 monitored by Kundi\'{c} et al. (1997). 

One might think that the flux variation of this macro-lensed 
quasar may be largely affected by microlensing events. 
Fortunately, however, 
most of the variability in this quasar 
is intrinsic one (see Figure 4 of Kundi\'{c} et al. 1997). 
Then, we adopt these light curves for testing models of 
intrinsic variability of AGNs.
%Note also that 0957+561 is a radio
%loud quasar and thus not an entirely fair test of the SB model.  
%(In this context, we do not preclude the presence of a black hole 
%in our SB model,
%since otherwise radio emission of 0957+561 cannot be accounted for.)
%Nevertheless,
%we confine the present analysis to this best available data set but
%note that no firm conclusions can be achieved until good quality light
%curves are available for a statistically valid sample of AGN.

Instead of PSD, we use a structure function [$V(\tau)$] analysis, 
which is almost equivalent to PSD analysis but suitable for 
gapped data, as is often the case in AGN light curves.
In short, SF expresses a curve of 
growth of variability with time-lag, 
and when time series of magnitude 
[$m(t_i), i = 1, 2, \ldots ; t_i < t_j $] is given, 
it is defined as 
\begin{eqnarray}
 V(\tau) \equiv \frac{1}{N(\tau)}
  \sum_{i<j}\left[\, m(t_i) \, - \, m(t_j) \,\right]^2, 
\end{eqnarray}
where summation is made over all pairs in which $t_j - t_i = \tau$, 
and $N(\tau)$ denotes a number of such pairs.

Usually, SF shows a power-law portion at smaller time-lags, 
\begin{eqnarray}
[V(\tau)]^{1/2} \propto \tau^{\beta}. 
\end{eqnarray}
This index $\beta$ also contains
important informations concerning the variability mechanism, 
like a power-law index of PSD. 
In fact, their indices are related each other by 
$\alpha = 1 + 2\beta$ in the limit of infinite and continuous data points.

We analyzed these light curves of Q0957+561 and showed that 
the power-law index $\beta$ is about 0.35 (Kawaguchi et al. 1998), 
which is consistent with 
the value known for $\alpha$ (1.5 - 2.0). 
Although we can not see the turn-off in the SF, 
which corresponds to the variability timescale, 
we expect to find it around 2 - 3 years, 
as observed in other quasars (Hook et al. 1994), 
if we obtain longer light curve.

\subsection{Disk-Instability Model} \label{DI}

\subsubsection{\large \it Procedure and Results \\}

The first model we are discussing is the disk instability model, 
whose original idea was proposed by Bak et al. (1988) as a sand-pile model 
to explain fluctuation properties of complex systems, such as earthquakes. 
They considered random input of sand grains onto a sand pile, 
and imposed a certain rule; 
when a slope at some site exceeds the critical value, 
an avalanche occurs, and the slope decreases below the critical value.
Then, they found that such a system evolves to and stays at a
self-organized critical (SOC) state, in which any size of avalanche
flows can occur, and thus producing $1/f$-like fluctuations.

In fact, this SOC model has been modified to the case of black-hole
accretion flows and applied to X-ray variability 
of Cyg X-1 (Mineshige et al. 1994; Takeuchi et al. 1995). 
We get a great success in reproducing the basic shapes of PSD
and the smooth shot-size distribution. 

Here, we propose
that an accretion disk (or flow) of AGN also 
stays at an SOC state, and calculate
flux variations expected by the model.
There are two key assumptions.
The first is that the disk is locally unstable. 
Probably, the local instability is of magnetic origin, 
leading to magnetic reconnection (Matsumoto et al. 1998).
% such as Balbus-Hawley instability (Matsumoto et al. 1998).
The second assumption is that each site of the disk interacts 
each other via avalanche. 
\begin{figure}
%\epsscale{0.3}
\plotfiddle{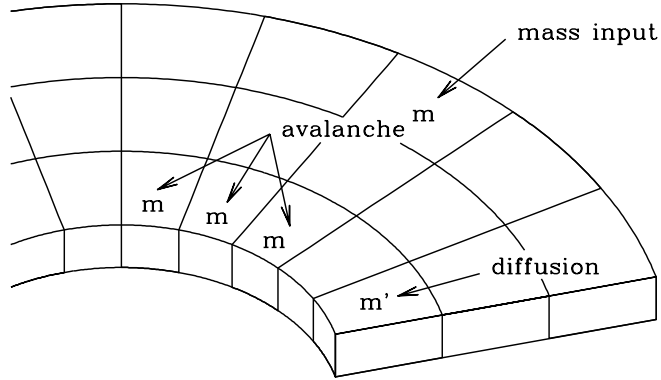}{45mm}{0}{55}{55}{-160}{-195}
% \vspace{52mm}
\caption{A schematic view of our cellular automaton model.} 
\end{figure}

The procedure of our cellular-automaton simulation is as follows 
(Figure 2). 

(i) First, we divide the disk into numerous cells. 
Then, we assume each cell behaves as a reservoir, 
which is quiescent until the mass density exceeds the critical value. 
In a word, even when mass input is steady, 
output will be episodic. 

(ii) We put one mass particle ($m$) at the outermost ring, 
which represents a mass supply to the disk.

(iii) Then, for unstable cells, 
where mass density exceeds a critical value, 
we set an avalanche flow. 
In other words, we move three mass ($3 m$) 
particles into inner adjacent cells. 

(iv) Aside from such critical behavior, 
we take account into viscous diffusion ($m'$ at each annulus). 
In general, effect of the diffusion is much less than 
avalanche flow. 

(v) We repeat the above procedures 
and we can draw the resultant light curve, assuming 
that the radiative energy is proportional to the potential energy released. 

We calculated models for some parameter sets, finding
that the resultant SFs exhibit 
power-law indices ($\beta$) of 0.4 - 0.5, 
which are close to the observed value ($\sim 0.35$).
Here, we should remark on the robustness of the cellular-automaton rule.
The results are quite insensitive to the detailed rules,
such as the number of cells, number of fallen particles, 
and prescription of the critical mass etc.
The only influential parameters are the ratio of the diffusion flow
over avalanche flow, $m^\prime/m$, and the size of the disk.
%The former affects the power-law index of PSD, while the latter
%determines the variability timescale (or break frequency).

\subsubsection{\large \it What Determines the Power-Law Index and 
Variability Timescale? \\}

\begin{figure}
\plotfiddle{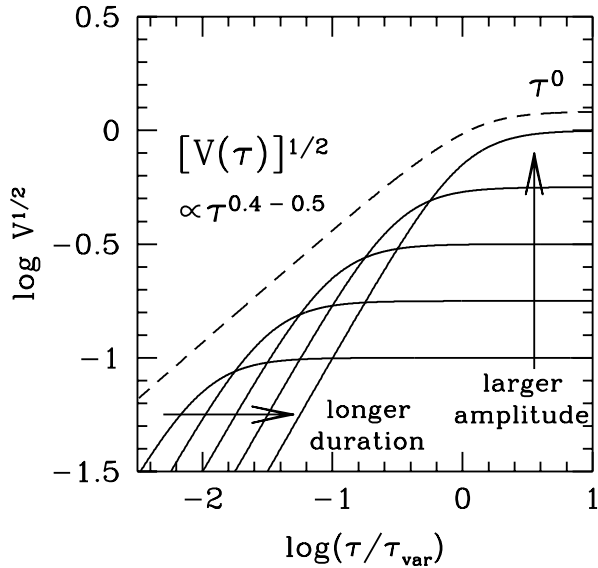}{70mm}{0}{70}{70}{-215}{-195}
\caption{Typical structure function expected by the DI model (dashed line) 
and structure functions of individual flares (solid lines).} 
\end{figure}
Figure 3 illustrates what determines the power-law index and 
variability timescale in the DI model. 
Dashed line represents a typical SF; 
a gradual ($\beta \sim$ 0.4 - 0.5) power-law increase 
at smaller time lags and a plateau 
over a variability timescale ($\tau_{var}$). 
The solid lines show SFs of individual flares. 
In general, smooth time-symmetric flares produce steeper SFs 
than $\tau^{0.5}$, as displayed by those solid lines. 
Thus, in order to explain the observed gradual slope, 0.35, 
there must be numerous shots with a wide variety of 
shot-amplitude and -duration.
More, smaller-amplitude and shorter-duration flares should
occur more frequently than larger and longer ones.

Thus, the index does not sensitively depend on the shot profile.
Instead, the distribution of shot-amplitude and -duration is more important.
Therefore, the index will not be largely changed even when the light curves of 
individual flares are modified. 

According to this picture,
the variability timescale is determined by the duration of the
largest flare.  
If fluctuating part of the disk is advection-dominated,
that duration will be 
of the order of the free-fall timescale from the outer rim.
Then, the variability timescales of several hundreds days roughly 
corresponds to size of 100 Schwarzschild radii ($r_{\rm g}$)
for a black hole mass of $10^8 M_\odot$;
\begin{equation}
 \tau_{\rm var} \approx \frac{r}{v_{\rm r}} 
    = \ 160 \left(\frac{v_{\rm r}}{0.1v_{\rm ff}}\right)^{-1}
	  \left(\frac{r}{10^2r_{\rm g}}\right)^{3/2}
          \left(\frac{M}{10^8M_\odot}\right)~{\rm day},
\end{equation}
where $v_{\rm r}$ and $v_{\rm ff}$ are radial velocity and free fall velocity, 
respectively.

\subsection{Starburst Model} \label{SB}

\subsubsection{\large \it Procedure and Results \\}

As an alternative model for radio-quiet AGNs, 
the starburst model has been investigated by Prof. Terlevich 
and his collaborators (e.g. Terlevich et al. 1992). 
According to the starburst model, optical variability 
is explained as superposition of supernova (SN) explosions. 
In the superposition, the released energy and duration 
of each SN explosion are varied within a factor of 2 in a Gaussian way around 
the mean values given a priori. 
As the supernova rate increases, resultant light curves are 
less variable in magnitudes. 
In other words, luminous AGNs must be less variable, 
which roughly agrees with the observed trend 
(Cristiani et al. 1996; see also Paltani \& Courvoisier 1997). 

We followed Aretxaga et al. (1997) and calculated light curves finding that 
the power-law indices $\beta$ expected from the model are around 0.7 - 0.9, 
which is significantly larger than the observed value, 0.35. 
Although the quasar we used here for the comparison is a radio-loud object 
and this may not be a fair test for the starburst model, 
we assume that optical variability of radio-loud objects is not very 
different from that of radio-quiet ones.

\subsubsection{\large \it What Determines the Power-Law Index and 
Variability Timescale? \\}

Then, why are the results so different between the two models?
In both models, something like shots or flare-like events 
are superposed almost randomly in the light curves. 
The critical point is that in the SB model 
there will not be large variation in the shot-amplitude and -duration 
over many orders of magnitude; 
timescales and released energies of each explosion can vary 
only by some factor. 
This is the substantial difference from the disk instability model. 

As the result, both the timescale and power-law index of SF 
are determined by a typical light curve of SN explosions, 
as shown in Figure 4. 
\begin{figure}
\plotfiddle{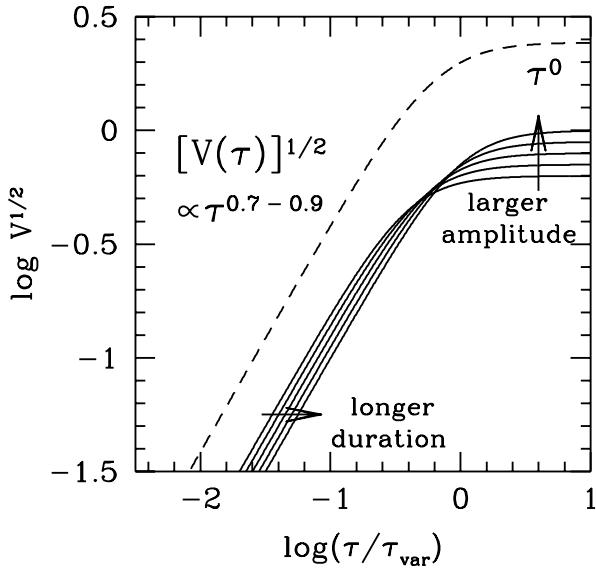}{70mm}{0}{70}{70}{-215}{-195}
\caption{Typical structure function expected by the SB model (dashed line) 
and structure functions of individual explosions (solid lines).} 
\end{figure}

To produce gradual index in the SF, 
we need much shorter events; for instance, 
SN explosions decaying over days. 
This is probably not the case in actual situation. 
It is possible that a thermal instability within 
SN remnants may change the typical light curve to be more 
fluctuating over short timescales (Cid Fernandes et al. 1996), 
so that the resultant power-law index 
will be consistent with the observed one.

\section{Other Issues}

We have focused on AGN light curve itself, 
with structure function analysis. 
In this section, we will discuss several related issues, 
especially the wavelength dependency of the variability.

First, there is a clear anti-correlation between fractional amplitudes and 
observed wavelength (e.g. Di Clemente et al. 1996). 
In other words, flux at shorter wavelength 
varies with larger amplitudes, compared to that at longer wavelength. 
It is also known that optical/UV 
continuum becomes ``harder'' as it gets brighter 
(Maoz et al. 1993; Peterson et al. 1994; Tr\`{e}vese 1998). 
These properties should contain important clues to the understanding
of the physics and site of variability generation. 
However in dealing with the spectral variability, 
we must be careful about the mixture of numerous emission lines and 
Balmer continuum, and also contaminations of other components. 

Next, the flux variations at different wave-bands are almost simultaneous
(e.g. Edelson et al. 1996), 
but recently it is reported that
there seems to exist some lags by several days or less than a day; 
(i) Optical variations tend to lag behind UV variations by $\sim$ 1 day 
(Peterson et al. 1998; Courvoisier 1998)
(ii) Flux variations of NGC 7469 in UV band sometimes 
lead X-ray ones by $\sim$ 4 days and sometimes they are simultaneous 
(Nandra et al. 1998).

In tentative pictures for central engines of AGNs, such as 
a cold accretion disk with hot corona
or cold clouds within hot corona, 
we had expected that X-ray variations lead Opt./UV variations, if we could 
detect such a small lags.
The new results of Nandra et al. (1998) are 
challenging to the tentative pictures. 
% Thus, we have now several useful results of observations for constructing 
Thus, it is a good time to construct 
a wavelength-dependent model of AGN variability 
including an anti-correlation of wavelength -- variation amplitude 
and spectral variability.
%it is about time to construct a wavelength-dependent model 
%including a wavelength -- variation amplitude anti-correlation 
%and spectral variability. 

\section{Summary}

\hspace*{\parindent}(1) As far as the power-law index in structure functions 
is concerned, the disk instability model is favored, 
since this model can produce a wide variety of 
shot-amplitudes and -durations, 
% in their light curves, 
as found in observed light curves.

(2) Observed variability timescales and power-law indices are the 
key properties to testing models.

(3) Constructing a wavelength and time dependent model of AGNs 
remains as a future work.

\acknowledgments

One of the authors (TK) is grateful to Prof. Edward Khachikian 
for the invitation to this stimulating conference. 
He also thanks to the Yamada Science Foundation for financial support.


\begin{references}
\reference Aretxaga, I., Cid Fernandes, R., \& Terlevich R. J. 1997, 
	\mnras, 286, 271
\reference Bak, P., Tang C., \& Wiesenfeld K.  1988, \pra, 38, 364
\reference Cid Fernandes, R. et al. 1996, \mnras, 283, 419
\reference Clavel, J. C. et al. 1991, \apj, 366, 64
\reference Courvoisier T. J.-L. 1998, these proceedings
\reference Cristiani, S. et al. 1996, \aap, 306, 395
\reference Di Clemente A. et al. 1996, \apj, 463, 466
\reference Edelson, R. A. et al. 1996, \apj, 470, 364
\reference Fahlman, G. G., \& Ulrych T. J.  1975, \apj, 201, 277
\reference Hayashida, K. et al. 1998, \apj, 500, 642
\reference Hook, I. M. et al. \mnras, 268, 305
\reference Kawaguchi, T., Mineshige S., Umemura M., \& Turner E. L.  1998, 
	\apj, 504, 671
\reference Korista, K. T. et al.  1995, \apjs, 97, 285
\reference Kundi\'{c}, T. et al. 1997, \apj, 482, 75
\reference Leighly, K. M., \& O'Brien, P. T. 1997, \apjl, 481, L15
\reference Maoz D. et al. 1993, \apj, 404, 576
\reference Matsumoto, R., \& Shibata, K. 1998, in preparation
\reference Mineshige, S., Takeuchi, M., \& Nishimori, H. 1994, \apjl, 435, L125
\reference Nandra K. et al. 1998, \apj, 505, 594
\reference Paltani S., \& Courvoisier T. J.-L. 1997, \aap, 323, 717
\reference Peterson, B. M. et al. 1994, \apj, 425, 622
\reference Peterson, B. M. et al. 1998, \pasp, 110, 660
\reference Takeuchi, M., Mineshige, S., \& Negoro, H. 1995, \pasj, 47, 617
\reference Terlevich, R. et al. \mnras, 255, 713
\reference Tr\`{e}vese, D. 1998, these proceedings
\reference Ulrich, M.-H., Maraschi L., \& Urry, C. M.  1997, \araa, 35, 445
\end{references}
\end{document}